\documentclass[aps,amsmath,twocolumn,prl,showpacs]{revtex4}

\usepackage{graphicx}
\usepackage{color}

\begin{document}

\preprint{}

\title{Polariton parametric photoluminescence in spatially inhomogeneous systems}

\author{Davide Sarchi}
\email[]{davide.sarchi@epfl.ch}
\affiliation{Institute of Theoretical Physics, Ecole Polytechnique F\'ed\'erale de Lausanne EPFL, CH-1015 Lausanne, Switzerland}
\author{Michiel Wouters}
\affiliation{Institute of Theoretical Physics, Ecole Polytechnique F\'ed\'erale de Lausanne EPFL, CH-1015 Lausanne, Switzerland}
\author{Vincenzo Savona}
\affiliation{Institute of Theoretical Physics, Ecole Polytechnique F\'ed\'erale de Lausanne EPFL, CH-1015 Lausanne, Switzerland}

\date{\today}

\begin{abstract}
A general theory of polariton parametric photoluminescence in spatially inhomogeneous systems is developed. The quantum Langevin equations are solved in a generalized Bogoliubov de Gennes approximation.
We apply the formalism to the specific case of a disordered microcavity. In this case, we numerically solve the equations for the coherent emission and the photoluminescence. We describe the effect of the exciton and photon disorder on the photoluminescence pattern exhibited in momentum space, finding a good agreement with the experimental observations.
\end{abstract}


\maketitle                   

Polariton photoluminescence is the paradigm for generation of quantum correlated states of electronic excitations in semiconductor devices. A pair of quantum correlated polaritons are generated via the scattering process $2 k_p\rightarrow k_s+k_i$, i.e. two polaritons in the pumped mode $k_p$ are scattered into one polariton in the ``signal'' mode $k_s$ and one polariton in the ``idler'' mode $k_i$ \cite{ciuti03}. Since all the k-modes satisfying the energy-momentum conservation rules are admitted as final modes, the resulting two-polariton state is an entangled state.
Experimental evidence of the quantum nature of the correlations between the signal and the idler polaritons has been reported \cite{langbeinprl,giacobino07}. Polaritons are also advocated as a vector of quantum correlations, in particular allowing the generation and the control of long-range quantum correlations that are stored in other systems, for example providing the ideal system for realizing spin coupling within a very long range \cite{piermarocchi06}. This possible application is very promising,
because of the recent engineering of new microstructures able to trap polaritons in spatially inhomogeneous regions \cite{kaitouni06,bajoni07}. These polariton traps are the ideal candidates for designing polariton quantum devices.
This scenario calls for a general theory of parametric photoluminescence in presence of spatial inhomogeneity.
In addition, the important role of structural disorder has been highlighted \cite{whittaker98,sanvitto06,savona07}, in particular in relation to parametric photoluminescence \cite{whittaker98,savona07}. It has been shown that the photoluminescence from idler modes is suppressed because it is very sensitive to disorder on the exciton component.

In this letter, we develop a general formalism for describing parametric photoluminescence in presence of arbitrary spatial inhomogeneity. The theory is based on the Bogoliubov de Gennes approach, typically used for describing Bose-Einstein condensation (BEC) in inhomogeneous systems. Bogoliubov theory assumes the presence of a large coherent field, that is treated classically, resulting in a linearized Hamiltonian for the quantum fluctuation field. In the present case, the classical Bogoliubov field is induced by the pump, while the residual quantum fluctuation field describes the actual parametric photoluminescence. Our theory is formulated in the basis of the exciton and photon fields and includes the linear exciton-photon coupling and the typical exciton non-linearities \cite{rochat00}. After deriving the general theory, we present the predictions for a disordered system and discuss the comparison with experiments \cite{langbein04prb}.

We treat the exciton and the photon fields $\hat{\Psi}_{x(c)}({\bf r})$ as Bose fields \footnote{We consider here a single polarization state \cite{carusotto05}, but the theory can be generalized to include the vector nature of the fields \cite{shelykh06}.} and we consider the effective Hamiltonian \cite{rochat00,bentabou01}
\begin{equation}
\hat{H}=\hat{H}_{0}+\hat{H}_{R}+\hat{H}_{x}+\hat{H}_{s}, \label{eq:Hcomp}
\end{equation}
where
$\hat{H}_{0}=\int d{\bf r}\hat{\Psi}^{\dagger}_{x}({\bf r})[-(\hbar^2/2m_x)\nabla^2+U_{x}({\bf r})]\hat{\Psi}_{x}({\bf r})+
\int d{\bf r}\hat{\Psi}^{\dagger}_{c}({\bf r})[\epsilon_c(\vec{\nabla})+U_{c}({\bf r})]\hat{\Psi}_{c}({\bf r})$
is the unperturbed term,
$\hat{H}_{R}=\Omega_{R}\int d{\bf r}[\hat{\Psi}^{\dagger}_{x}({\bf r})\hat{\Psi}_{c}({\bf r})+h.c.]$
is the term describing the exciton-photon coupling,
$\hat{H}_{x}=\frac{1}{2}v_x\int d{\bf r}\hat{\Psi}^{\dagger}_{x}({\bf r})\hat{\Psi}^{\dagger}_{x}({\bf r})\hat{\Psi}_{x}({\bf r})\hat{\Psi}_{x}({\bf r})$
is the effective 2-body exciton interaction term,
modeling both Coulomb interaction and the effect of Pauli
exclusion on electrons and holes \cite{rochat00}, and
$\hat{H}_{s}=v_{s}\int d{\bf r}[\hat{\Psi}^{\dagger}_{c}({\bf r})\hat{\Psi}^{\dagger}_{x}({\bf r})\hat{\Psi}_{x}({\bf r})\hat{\Psi}_{x}({\bf r})+h.c.]$ is the term modeling the saturation of the exciton oscillator strength \cite{rochat00}. Here we neglect the momentum dependence of the scattering matrix elements $v_x$ and $v_s$, because for our present purposes we will consider only scattering processes between states with small momentum.

We consider the system evolving under a continuous monochromatic optical pump, $F({\bf r},t)= e^{-i\omega_p t}F^0({\bf r})$, and we assume that the exciton and photon fields decay in time with the rates $\gamma_{x}$ and $\gamma_c$, respectively. In this non-equilibrium regime, the two fields are written as \cite{carusotto05}
\begin{equation}
\hat{\Psi}_{x(c)}({\bf r},t)=e^{-i\omega_p t}\left[\Phi_{x(c)}({\bf r})+\tilde{\psi}_{x(c)}({\bf r},t)\right],
\label{eq:bogans}
\end{equation}
i.e. as the sum of a classical term $\langle \hat{\Psi}_{x(c)}({\bf r},t)\rangle=e^{-i\omega_p t}\Phi_{x(c)}({\bf r})$,
describing the coherent field generated by the pump and evolving accordingly with the pump frequency $\omega_p$, and a Bose field
$\tilde{\psi}_{x(c)}({\bf r},t)$, describing the fluctuations. By averaging the Heisenberg equations of
motion of $\hat{\Psi}_{x(c)}({\bf r},t)$, we obtain the two coupled equations
\begin{eqnarray}
\hbar\omega_p\Phi_x({\bf r})&=&\left(-\frac{\hbar^2\nabla^2}{2 m_x}+U_{x}({\bf r})-i\gamma_x+v_x\mid \Phi_x({\bf r})\mid^2\right.\nonumber\\
&+&\left.+2v_s\mbox{Re}\left\{\Phi_x^*({\bf r})\Phi_c({\bf r})\right\}\right)\Phi_x({\bf r})\nonumber\\
&+&\left(\Omega_R+v_s\mid \Phi_x({\bf r})\mid^2\right)\Phi_c({\bf r}), \label{eq:GP_X}\\
\hbar\omega_p\Phi_c({\bf r})&=&\left[\epsilon_c(\vec{\nabla})+U_{c}({\bf r})-i\gamma_c\right]\Phi_c({\bf r})\nonumber\\
&+&\left(\Omega_R+v_s\mid \Phi_x({\bf r})\mid^2\right)\Phi_x({\bf r})+F^0({\bf r})\,, \label{eq:GP_C}
\end{eqnarray}
defining the spatial shape $\Phi_{x,c}({\bf r})$ of the coherent fields.

By adopting an input-output formalism \cite{carusotto06} and by linearizing the time evolution of the excitation field, we derive the quantum Langevin equation for the fluctuations \cite{verger07}
\begin{equation}
i\hbar\partial_t{\bf \tilde{\Psi}}({\bf r},t)=\hat{M}{\bf \tilde{\Psi}}({\bf r},t)+{\bf \tilde{f}}({\bf r},t)\,, \label{eq:QLang}
\end{equation}
where ${\bf \tilde{\Psi}}=(\tilde{\psi}_x, \tilde{\psi}_x^{\dagger}, \tilde{\psi}_c, \tilde{\psi}_c^{\dagger})^T$ is the four-component fluctuation field, while
${\bf \tilde{f}}=(\tilde{f}_x, \tilde{f}_x^{\dagger}, \tilde{f}_c, \tilde{f}_c^{\dagger}$) defines the fluctuation field, with correlations $\langle\tilde {f}_{\xi}({\bf r},t)\rangle_{\bf f}=0$ and $\langle\tilde{f}_{\xi}({\bf r},t) \tilde{f}^{\dagger}_{\chi}({\bf r}^{\prime},t')\rangle_{\bf f}=2\pi \gamma_{\chi}\delta_{\xi,\chi}\delta({\bf r}-{\bf r}^{\prime})\delta(t-t')$ \cite{verger07}.
The matrix $\hat{M}$ has the Bogoliubov de Gennes form \cite{fetter72}
\begin{equation}
\hat{M}=\left(\begin{array}{cccc}
\hat{T}_x-i\gamma_x & \Sigma^{xx}_{12} & \tilde{\Omega}_R & \Sigma^{xc}_{12}\\
-(\Sigma^{xx}_{12})^* & -\hat{T}_x-i\gamma_x & -(\Sigma^{xc}_{12})^* & -\tilde{\Omega}_R \\
\tilde{\Omega}_R & \Sigma^{xc}_{12} & \hat{T}_c-i\gamma_c & 0 \\
-(\Sigma^{xc}_{12})^* & -\tilde{\Omega}_R & 0 & -\hat{T}_c-i\gamma_c\end{array}\right)\,, \label{eq:MBdG}
\end{equation}
where
\begin{eqnarray}
\hat{T}_x({\bf r})&=&-\frac{\hbar^2\nabla^2}{2 m_x}+U_{x}({\bf r})-\hbar\omega_p+2v_x\mid \Phi_x({\bf r})\mid^2\nonumber\\
&+&4v_s\mbox{Re}\{\Phi_x^*({\bf r})\Phi_c({\bf r})\}\,, \\
\Sigma_{12}^{xx}({\bf r})&=&v_x\Phi_x^2({\bf r})+2v_s\Phi_x({\bf r})\Phi_c({\bf r})\,, \\
\tilde{\Omega}_R({\bf r})&=&\Omega_R+2v_s\mid \Phi_x({\bf r})\mid^2\,,\\
\Sigma^{xc}_{12}({\bf r})&=&v_s\Phi^2_x({\bf r})\,.\label{eq:MBdG2}
\end{eqnarray}
We notice that the interaction terms in Eqs. (\ref{eq:GP_X}), (\ref{eq:GP_C}) and (\ref{eq:MBdG}) are written within the Bogoliubov limit, i.e. they only depend on the coherent fields, because we have assumed that the incoherent populations are vanishingly small. This is always true in the regime of parametric photoluminescence that we are considering in the present work. However, in the same spirit as for symmetry breaking theories of BEC \cite{shi98}, our symmetry breaking approach allows to systematically include the higher order interaction processes involving excitations, and thus to treat a regime of higher density \cite{ciuti03,sarchi08}.

By Fourier transforming Eq. (\ref{eq:QLang}) in the frequency domain
\begin{equation}
\hbar\omega{\bf \tilde{\Psi}}({\bf r},\omega)=\hat{M}{\bf \tilde{\Psi}}({\bf r},\omega)+{\bf \tilde{f}}({\bf r},\omega)\,, \label{eq:QLang_omega}
\end{equation}
we have access to the frequency resolved exciton (photon) spatial density $n_{x(c)}({\bf r},\omega) =\langle\tilde{\psi}_{x(c)}^{\dagger}({\bf r},\omega)\tilde{\psi}_{x(c)}({\bf r},\omega) \rangle$. By using the correlation properties of the fluctuation field and after some algebra, we can write the densities in the useful form
\begin{equation}
n_{x(c)}({\bf r},\omega)=\sum_{l=1}^{4}\int d{\bf r'}\left|\langle{\bf r'},l|\hat{M}_{\omega}^{-1}| {\bf r},1(3)\rangle\right|^2 \Gamma_l\,, \label{eq:PLres}
\end{equation}
where ${\bf \Gamma}=(0, 2\pi \gamma_{x},0,2\pi \gamma_{c})^T$ and $\hat{M}_{\omega}=\hat{M}-\hbar\omega{\bf 1}$ is evaluated in the basis $|({\bf r},j)\rangle$, where $|{\bf r}\rangle$ spans the 2D real space, while the label $j=1,...,4$ refers to the bloc form of $\hat{M}$, Eq.~(\ref{eq:MBdG}). In this work, we focus on the regime of spontaneous photoluminescence and we neglect the density of excitations in the expression of the correlations $\Gamma_l$.

The excitation fields ${\psi}_j({\bf k},\omega)$ expressed in the in-plane momentum space obey an equation formally equivalent to Eq. (\ref{eq:QLang_omega}) and the densities in momentum space are given by
\begin{equation}
n_{x(c)}({\bf k},\omega)=\sum_{l=1}^{4}\int d{\bf k'}\left|\langle{\bf k'},l|\hat{M}_{\omega}^{-1}| {\bf k},1(3)\rangle\right|^2 \Gamma_l\,. \label{eq:PLres_k}
\end{equation}
Full information about the photoluminescence emission is obtained from
Eqs. (\ref{eq:PLres}) and (\ref{eq:PLres_k}), that can be numerically solved for any shape of the external potentials $U_{x,c}({\bf r})$ and of the exciting pump $F^0({\bf r})$. The form of Eqs. (\ref{eq:PLres}) and (\ref{eq:PLres_k}) is particularly advantageous for numerical computations, because of the efficiency of the algorithms for the treatment of sparse matrices.

We now apply the model to the specific situation in which the external potentials $U_{x,c}({\bf r})$ describe the structural disorder naturally present in the polariton system \cite{savona07}. We adopt parameters modeling a typical GaAs microcavity \cite{langbein04prb}. In particular, we assume radiative linewidths $\gamma^{rad}_x=0.06$ meV and $\gamma^{rad}_c=0.13$ meV, and we assume that the external pump have frequency $\hbar\omega_p=-1.2~\mbox{meV}$ with respect to the heavy-hole exciton ground state and in-plane momentum ${\bf k}_p=(1.73,0)~\mu\mbox{m}^{-1}$.
The photon disorder potential is assumed to be Gauss correlated in space \cite{savona07}
\begin{equation}
\langle U_{c}({\bf r}) U_{c}({\bf r'})\rangle=\sigma_{c}^2e^{-|{\bf r}-{\bf r'}|^2/\xi_{c}^2}\,,
\label{eq:dispot}
\end{equation}
with correlation amplitude $\sigma_c=0.1$ meV and correlation length $\xi_{c}=7~\mu\mbox{m}$. These values result into an inhomogeneous broadening $\Delta E_c\approx50~\mu\mbox{eV}$ for the cavity photon resonance. The exciton disorder is assumed with a short correlation (white noise) on our computation grid and its amplitude is fixed in order to give an inhomogeneous broadening $\Delta E_x\approx0.75$ meV for the exciton resonance \cite{savona07}.

\begin{figure}[ht]
\includegraphics[width=.45 \textwidth]{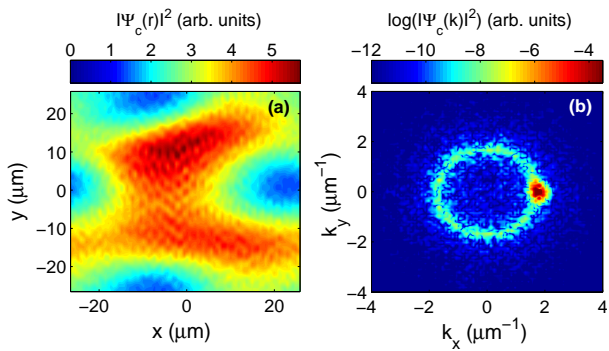}
\caption{(Color online) (a) Square modulus of the coherent field $|\Phi_C({\bf r})|^2$ in the real space, in linear color scale. (b) Square modulus of the coherent field $|\Phi_C({\bf k})|^2$ in the in-plane momentum space, in logarithmic color scale. Notice the spot at $(1.73,0)~\mu\mbox{m}^{-1}$ corresponding to the incident pump field (intensity off scale).} \label{fig1}
\end{figure}
The solution of Eqs. (\ref{eq:GP_X}) and (\ref{eq:GP_C}) is plotted in Fig. \ref{fig1}. In Fig. \ref{fig1}(a) we show the square modulus of the resulting coherent photon field $|\Phi_c({\bf r})|^2$ in real space. The photon disorder is responsible for the localization of the field over a distance of tens of microns, while the exciton disorder is responsible for the additional short-length fluctuations of the amplitude of the field.
The square modulus of the coherent photon field $|\Phi_c({\bf k})|^2$ in the ($k_x$,$k_y$)-space is shown in Fig. \ref{fig1}(b). Here, the typical elastic ring of Rayleigh scattered polaritons is clearly visible \cite{houdre00,gurioli01,langbein02}.

\begin{figure}[ht]
\includegraphics[width=.43 \textwidth]{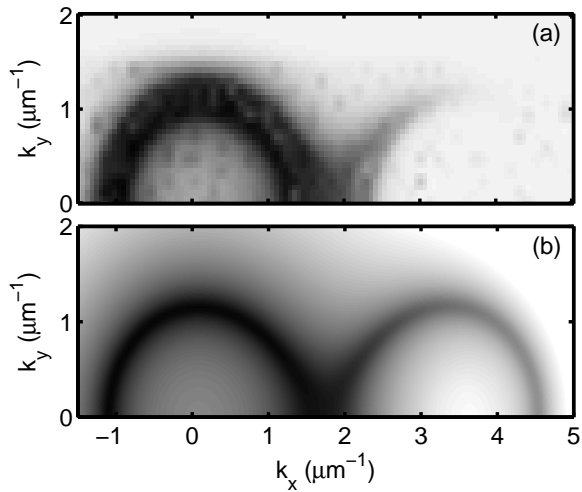}
\caption{Pattern in the momentum space of the photoluminescence for the disordered microcavity (a) and for a uniform microcavity (b), plotted in logarithmic color map.} \label{fig:2}
\end{figure}
We then solve Eq. (\ref{eq:PLres_k}) and compute, in the momentum space, the parametric photoluminescence intensity, which corresponds to the frequency-integrated photon density $I_{PL}({\bf k})=\int d\omega n_c({\bf k},\omega)$. In Fig. \ref{fig:2} we compare the resulting photoluminescence intensity (in panel (a)) with the one obtained for the uniform system (in panel (b)), i.e. in the absence of photon and exciton disorder. In the uniform case, a eight-shaped photoluminescence pattern, due to the momentum selection rules (the width of the eight-shape is due to the finite linewidths), clearly appears. We call ``Signal'' emission that for $k_x<k_{p_x}$, while ``idler'' emission that for $k_x>k_{p_x}$. In the disordered case, Fig. \ref{fig:2} (a), the momentum selection rule is lifted and the pattern is broadened and fragmented. At large momenta, the effect is dramatic, resulting into a complete spreading out of the idler emission over the excitonic states. The vanishing of the idler resonance is due to the strong effect of the exciton disorder in the large-momenta region of the polariton dispersion, where the very large effective mass makes the states sensitive to the disorder-induced broadening of the exciton resonance. This result agrees with the experimental observations, because a clear evidence of the eight-pattern has been reported only for a sample specially designed to minimize exciton disorder \cite{langbein04prb}. As expected from the linear response theory, at small momenta the effect of exciton disorder is less evident \cite{savona07}.
\begin{figure}[ht]
\includegraphics[width=.43 \textwidth]{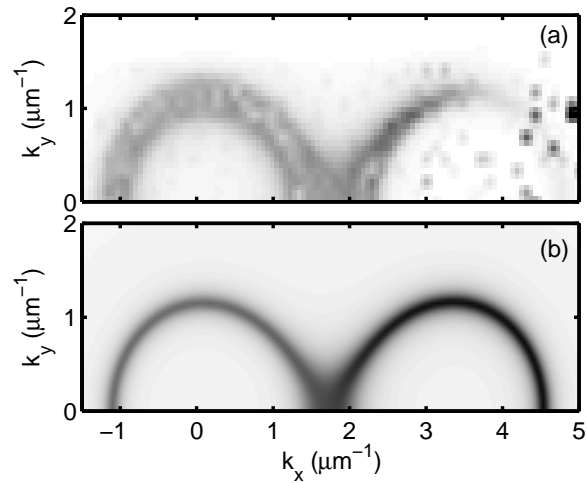}
\caption{Polariton population $n_p(k_x,k_y)$ in presence of disorder (a) and for a uniform system (b). The color map is in linear grey scale.} \label{fig:3}
\end{figure}
However, in Fig. \ref{fig:2}(a) we also notice the presence of speckles, both in the idler and in the signal emission. The qualitative difference between their distribution and the one of the speckles in the linear response (Fig. \ref{fig1}) is probably related to the non-linear nature of the parametric process.

Although speckles are hardly visible in photoluminescence, they are an indication of the dramatic effect of the exciton disorder on the polariton population $n_{p}({\bf k},\omega)=n_{c}({\bf k},\omega)+n_{x}({\bf k},\omega)$, i.e. the quantity which is more relevant for future applications in the generation of quantum correlated pairs of polaritons. This quantity is considered in Fig. \ref{fig:3}, where we compare the resulting frequency-integrated polariton population (panel (a)) with the one obtained for the uniform system (panel (b)). The effect of disorder is even more striking. The spots in the idler region indicate the fragmentation in the momentum space of the population distribution. This fragmentation also reflects on the population distribution at small momenta, which is broadened with respect to the homogeneous case.
Notice that in the uniform case, because of the signal-idler symmetry, the relative amplitude of the frequency integrated populations of the signal and idler modes depends on the ratio between linewidths, via the simple relation $\gamma_{\bf k} n({\bf k})=\gamma_{2{\bf k}_p-{\bf k}} n(2{\bf k}_p-{\bf k})$. In the disordered case, because of the localization of the modes, this relation is no longer valid.
\begin{figure}[ht]
\includegraphics[width=.43 \textwidth]{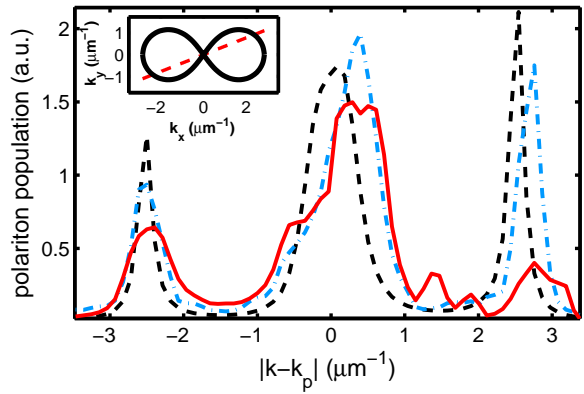}
\caption{Frequency integrated polariton population $n_p(|{\bf k}-{\bf k}_p|)$ along the cut shown in the inset. We compare the results obtained for a uniform system (dashed line), in presence of photon disorder only (dot-dashed) and in presence of both exciton and photon disorder (solid).} \label{fig:4}
\end{figure}

Fragmentation is even more evident in Fig. \ref{fig:4}, where we show the frequency integrated polariton population along one specific cut in the momentum plane (the dashed line shown in the inset of the figure). In the absence of disorder, the distribution is narrow around the signal and idler modes and the population is larger in the idler than in the signal, because of the longer radiative lifetime. When photon disorder is introduced, a broadening appears, corresponding to the softening of the momentum selection rule, but this effect is relatively small and the signal-idler pair is clearly visible.
In the presence of exciton disorder, the population in the idler region is significantly fragmented in momentum space. Correspondingly, the signal resonance is broadened. Indeed, the parametric process mixes the high energy exciton-like states and the low-momenta ones. Notice that the same effect can be seen in the idler when only photon disorder is present (dot-dashed line).
We have checked that, while the disappearance of the eight-shape of the photoluminescence pattern and the broadening of the population distribution in the signal can be qualitatively reproduced by introducing inhomogeneous linewidths \cite{whittaker98}, the fragmentation of the polariton population, due to the presence of localized states, can be predicted only by a full treatment of the spatial disorder.

In conclusion, we have developed a general formalism for describing the spontaneous parametric photoluminescence in a spatially inhomogeneous polariton system. The range of application of this formalism is very large, allowing the description of disordered systems and of the recently engineered polariton traps \cite{kaitouni06,bajoni07}. In the present work we have applied the formalism to study the appearance of the eight-shaped pattern in the photoluminescence of a disordered microcavity. We have shown that the disordered-induced fragmentation of the population distribution in the momentum space is responsible for a strong suppression and broadening of both the idler and the signal resonances. The important effect on the signal population is a consequence of its parametric coupling with the idler. Our formalism is the tool required for designing structures for the generation of quantum correlated polariton states.

We are grateful to I. Carusotto for enlightening discussions.

\bibliographystyle{apsrev}


\end{document}